\documentclass[sigconf]{acmart}
\usepackage{array}
\usepackage{float}
\usepackage{placeins}
\usepackage{caption}
\usepackage{subcaption}
\usepackage{caption}
\usepackage{subcaption}
\usepackage{float}
\usepackage{tikz}
\usepackage{mwe}
\usepackage{rotating}

\usepackage{bm}
\usepackage{bbm}
\usepackage{amsmath}
\usepackage{booktabs}
\usepackage{algorithm}
\usepackage{multirow}
\usepackage{enumitem}
\usepackage{makecell}
\usepackage{graphicx}

\AtBeginDocument{%
  \providecommand\BibTeX{{%
    \normalfont B\kern-0.5em{\scshape i\kern-0.25em b}\kern-0.8em\TeX}}}

\setcopyright{acmcopyright}
\copyrightyear{2023}
\acmYear{2023}
\setcopyright{acmlicensed}
\acmConference[SIGIR '23] {Proceedings of the 46th International ACM SIGIR Conference on Research and Development in Information Retrieval}{July 23--27, 2023}{Taipei, Taiwan.}
\acmBooktitle{Proceedings of the 46th International ACM SIGIR Conference on Research and Development in Information Retrieval (SIGIR '23), July 23--27, 2023, Taipei, Taiwan}
\acmPrice{15.00}
\acmISBN{978-1-4503-9408-6/23/07}
\acmDOI{10.1145/3539618.3591954}

\begin{document}

\title{Bayesian Knowledge-driven Critiquing with Indirect Evidence}

\author{Armin Toroghi}

\affiliation{
 \institution{University of Toronto}
 \city{Toronto}
 \state{ON}
 \country{Canada}}
 \email{armin.toroghi@mail.utoronto.ca}

\author{Griffin Floto}
\affiliation{%
 \institution{University of Toronto}
 \city{Toronto}
 \state{ON}
 \country{Canada}}
 \email{griffin.floto@mail.utoronto.ca}
\author{Zhenwei Tang}
\affiliation{
 \institution{University of Toronto}
 \city{Toronto}
 \state{ON}
 \country{Canada}}
 \email{josephtang@cs.toronto.edu}

\author{Scott Sanner}
\affiliation{
 \institution{University of Toronto}
 \city{Toronto}
 \state{ON}
 \country{Canada}}
 \email{ssanner@mie.utoronto.ca}
\authornote{Affiliate to Vector Institute of Artificial Intelligence, Toronto}

\renewcommand{\shortauthors}{Armin Toroghi, Griffin Floto, Zhenwei Tang, \& Scott Sanner}

\begin{abstract}
Conversational recommender systems (CRS) enhance the expressivity and personalization of recommendations through multiple turns of user-system interaction. Critiquing is a well-known paradigm for CRS that allows users to iteratively refine recommendations by providing feedback about attributes of recommended items.  While existing critiquing methodologies utilize direct attributes of items to address user requests such as \textit{‘I prefer Western movies’}, the opportunity of incorporating richer contextual and side information about items stored in Knowledge Graphs (KG) into the critiquing paradigm has been overlooked. Employing this substantial knowledge together with a well-established reasoning methodology paves the way for critique-based recommenders to allow for complex knowledge-based feedback (e.g., \textit{‘I like movies featuring war side effects on veterans’}) which may arise in natural user-system conversations. In this work, we aim to increase the flexibility of critique-based recommendation by integrating KGs and propose a novel Bayesian inference framework that enables reasoning with relational knowledge-based feedback. 
We study and formulate the framework considering a Gaussian likelihood and evaluate it on two well-known recommendation datasets with KGs. Our evaluations demonstrate the effectiveness of our framework in leveraging 
\emph{indirect} KG-based feedback (i.e., preferred relational properties of items rather than preferred items themselves), often improving personalized recommendations over a one-shot recommender by more than 15\%.
This work enables a new paradigm for using rich knowledge content and reasoning over indirect evidence as a mechanism for critiquing interactions with CRS.
\end{abstract}



\maketitle
\begin{figure}[t!]
  \centering
  \includegraphics[width=0.87\linewidth]{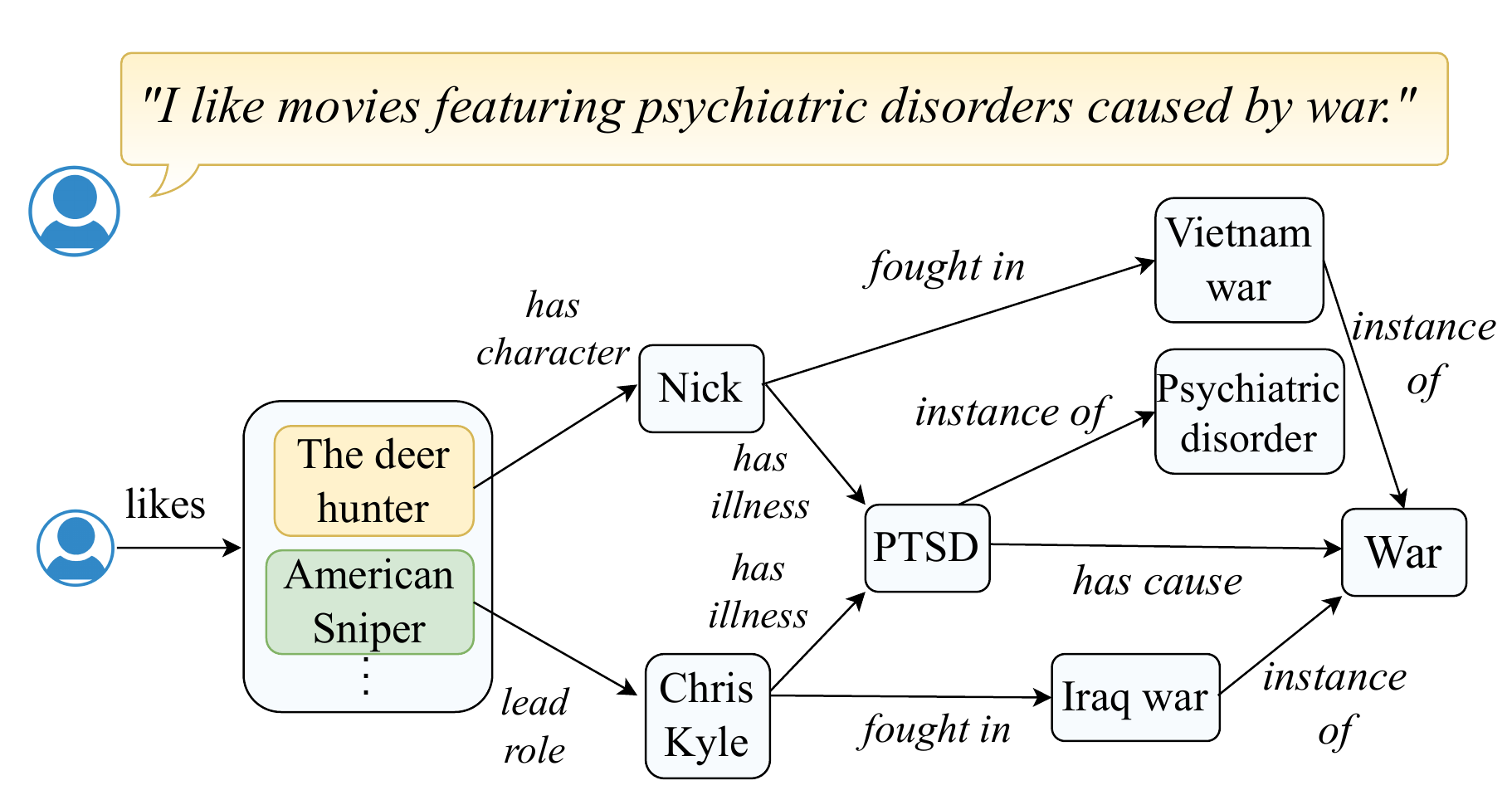}
  \caption{An example of knowledge-based feedback containing indirect evidence of user's interest and using the side information of knowledge graph to address it}
  \label{intro}
\end{figure}
\begin{figure*}[t]
\centering
\includegraphics[width=0.92\textwidth]
{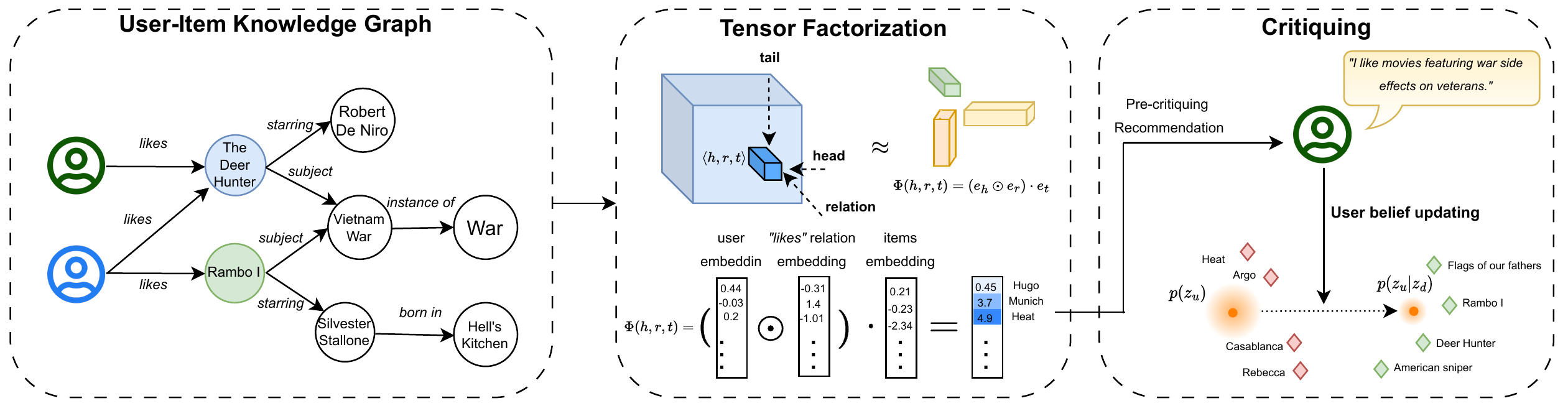}
\vspace{-8pt}
\begin{minipage}[t]{.375\linewidth}
\centering
\vspace{-18pt}
\subcaption{} \label{a}
\end{minipage}%
\hfill
\begin{minipage}[t]{.3\linewidth}
\centering
\vspace{-18pt}
\subcaption{}\label{b}
\end{minipage}
\begin{minipage}[t]{.32\linewidth}
\centering
\vspace{-18pt}
\subcaption{}\label{c}
\end{minipage}
\caption{General workflow of the BCIE framework. (a) user interaction data is combined with KG side information to create a user-item KG (b) A tensor factorization model is trained to learn entity and relation embeddings, which are then utilized to generate initial recommendations. (c) Indirect evidence of user’s interest is observed from the critique and conditioned on the embeddings of the indirect evidence ($z_d$), posterior update is performed on the user belief ($p(z_u)$) to refine it toward representations of items satisfying the critique and make refined recommendations in the next session.}
\label{workflow}
 \end{figure*}
\section{Introduction}

The growth of interest in recommender systems has given rise to demands and expectations for more expressive user-system interactions and as a response, Conversational Recommender Systems (CRS) have been introduced to facilitate a deeper understanding of a user's preferences through a sequence of user-system interactions \citep{2020jannach, sun2018conversational, crss1, crss2}. Critiquing is an evolving CRS paradigm that enables personalized recommendations by adapting to user's feedback in an iterative process \citep{chen2012critiquing}. In each critiquing session, the system recommends a set of items and in response, the user either accepts them or provides feedback about her preference toward a subset of recommended items or their attributes.
While previous studies have presented techniques to incorporate evidence of a user's interests in other items \citep{wrmf, sarwar2001item}, items from other domains \citep{sedhain2014social, rosli2015alleviating}, and item properties \citep{pazzani2007content}, the ability to personalize recommendations using knowledge-based feedback containing indirect evidence of the user's interests has not been addressed. An example of this type of feedback and the associated relational knowledge needed to address it are shown in Figure \ref{intro}. This type of feedback is referred to as \textit{indirect evidence} since the provided information does not target the user’s preferred items but a distribution of items that share those relational properties to different extents.

Addressing this type of feedback requires abundant side information about the users and/or items and their relational dependencies, and Knowledge Graphs (KG) as a rich source of side information have been widely used to assist in recommendation \citep{guo2020survey}. Despite the considerable attention devoted to KG-enhanced recommendation in recent years \citep{wang2018dkn,wang2018ripplenet,wang2019kgat}, most works in this realm have focused on enhancing recommendations in a one-shot paradigm, and the question of how KG side information could be utilized to address user’s feedback in a conversational setting has been understudied. Aligning KG-enhanced recommendation with a CRS requires a proper methodology for reasoning over the user's interests based on the KG information, which this work aims to address.

In this work, we make the following contributions: (i) We introduce
the Gaussian variant of a popular tensor factorization approach for KG-enhanced recommendation. (ii) We propose Bayesian Critiquing with Indirect Evidence (BCIE), a knowledge-driven critiquing framework, and formulate a Bayesian closed-form user belief updating methodology to enable critiquing CRSs to address indirect feedback.
(iii) We empirically show that BCIE results in considerable improvement of personalized recommendation over one-shot recommendation by evaluation on two 
datasets and a KG.

\section{Bayesian Critiquing with Indirect Evidence}
\label{sec:design_choices}

The overall workflow of the BCIE framework is shown in Figure \ref{workflow}. Here, we first define the problem of Bayesian knowledge-driven critiquing with indirect evidence, and next, illustrate how BCIE operates in the pre-critiquing and critiquing phases. 

\subsection{Problem Definition}
In the conversational critiquing problem setting that we investigate, denoting the set of available items by $I$, preference records of a user $u \in U$ are extracted from the user preference matrix $R \in \mathbb{R}^{|U| \times |I|}$ and transformed to a set of KG triples $H= \{ (u, likes, i) | i\in I\}$, that along with KG triples of item side information form a user-item KG as shown in Figure \ref{workflow}(a). By performing link prediction on the user-item KG, a set of items $R_{u}$ that seem pertinent to the user $u$ are recommended. Link prediction is performed using a scoring function $\Phi(\langle h,r,t \rangle)$ that estimates the plausibility of the existence of the relation $r$ from $h$ to $t$. Along with the recommended items, the set of facts $F=\{(h,r,t) | t \in R_{u} \lor h \in R_{u} \}$ related to the recommended items is presented to the user, and the user may either accept the recommendation or critique a fact $f\in F$ (e.g., the critique presented in Figure \ref{workflow}(c)). The recommender's duty in the next step is to update its belief in the user's interests and refine $R_{u}$ given $d_n$, the evidence of the user's taste observed from the critique at iteration $n$. Hence, the recommender needs a critique-modified recommendation function $f_m$, such that $\hat{R_u} = f_{m}(R_u, d_n)$.
This process continues either for a fixed number of iterations or until the user accepts the recommendation or leaves the conversation.

\subsection{Pre-critiquing phase} The initial stage of BCIE's workflow involves a standard knowledge-based recommendation problem, where the objective is to leverage side information of a KG to improve recommendations. We build our recommender upon SimplE, a well-known tensor factorization-based KG embedding model, because of its efficient computations and \emph{full-expressiveness}~\citep{kazemi2018simple}. This model assigns two embedding vectors $h_{e}$ and $t_{e}$ to each entity $e$ and two vectors $v_{r}$ and $v_{r^{-1}}$ to each relation $r$, and defines its scoring function for a triple $(e_{i},r,e_{j})$ as $\Phi(e_{i},r,e_{j})=\frac{1}{2}(\langle h_{e,i},v_{r},t_{e,j}\rangle + \langle h_{e,j},v_{r}^{-1},t_{e,i}\rangle)$, in which $\langle v,w,x \rangle =(v \odot w) \cdot x$ where $\odot$ is element-wise and $\cdot$ represents dot product. The model parameters are trained by forming a batch of size $N$ from a number of KG triples and negative samples and solving the following optimization problem:
\begin{equation}
    \label{simpleloss}
    \min_{\theta} \sum_{n=1}^{N} \log(1+ \exp(-y_{n}. \Phi(e_{i},r,e_{j})_{n}) + \lambda \| \theta \|_{2}^{2}
\end{equation}
in which $y\in \{+1,-1\}$ denotes the triple label, $\theta$ stands for the model parameters, $n$ is the iterator of batch samples, and $\lambda$ is the regularization hyperparameter. Using the learned embeddings of entities and relations, the set of items yielding the highest plausibility scores for $(user, likes, item)$ triples are picked for recommendation.

\subsection{Critiquing phase: Bayesian User Belief Updating with Indirect Evidence} 
In each critiquing session, the user provides knowledge-based feedback containing \textit{indirect} evidence of her preference. Notably, the type of critique studied in this work is not limited to being directly attributed to a particular item, rather, the user may provide indirect feedback about various relational attributes from different items that she likes. Hence, in the BCIE framework, we need to consider a distribution over representations of items that cover the user's interest, which is denoted by $\boldsymbol{z_m}$. To this end, 
we require to maintain a belief state over the user preferences, hereafter called \textit{user belief}, which is initially centered at the learned embedding of the user entity and update it conditioned on the user critiques. By exponentiating SimplE's loss function in Equation \ref{simpleloss} to form the maximum likelihood problem, it could be shown that the SimplE transforms the link prediction problem to a tensor factorized view of logistic regression by maximizing the sigmoid likelihood. However, in BCIE, we introduce a new variant of SimplE in which the sigmoid likelihood is replaced by a Gaussian, forming a conjugate pair with the Gaussian prior assumed over the user belief; this, in turn, enables tractable closed-form belief updates using Gaussian Belief Propagation (BP) \cite{weiss1999correctness}.


To update the user belief conditioned on the observed indirect evidence, we first need to update the item distribution. Assuming the evidence distribution as $\boldsymbol{z}_{d} \sim \mathcal{N}^{-1}(\boldsymbol{h_d}, \boldsymbol{J_d})$ and prior over item distribution as $\boldsymbol{z}_{m} \sim \mathcal{N}^{-1}(\boldsymbol{h_{m}}, \boldsymbol{J_{m}})$, where $\boldsymbol{h}$ and $\boldsymbol{J}$ are the potential vector and precision matrix of the corresponding distribution in the information form \citep{bishop2006pattern}, the posterior item distribution
becomes:
\begin{equation}
    p (\boldsymbol{z}_{m}|\boldsymbol{z}_{d}) \sim \mathcal{N}^{-1}(\boldsymbol{h_{d}} + \boldsymbol{h_{m}}, \boldsymbol{J_{d}} + \boldsymbol{J_{m}})
\end{equation}
At the next stage, we pass on this updated belief to the user distribution $\boldsymbol{z}_{u} \sim \mathcal{N}^{-1}(\boldsymbol{h_{u}}, \boldsymbol{J_{u}})$. First, we can write the joint distribution of $\boldsymbol{z}_{u}$ and $\boldsymbol{z}_{m}$ conditioned on the evidence as:
\begin{multline}
 p (\boldsymbol{z}_{u},\boldsymbol{z}_{m}|\boldsymbol{z_d}) \propto \\ \exp\left\{{-\frac{1}{2}}\boldsymbol{z}_{u}^{T} \boldsymbol{J}_{\boldsymbol{u}} \boldsymbol{z}_{u} + \boldsymbol{h_{u}}^{T} \boldsymbol{z}_{u} - \frac{1}{2}\boldsymbol{z}_{m}^{T} \boldsymbol{J}_{\boldsymbol{z}} \boldsymbol{z}_{m} + \boldsymbol{h}_{\boldsymbol{z}}^{T} \boldsymbol{z}_{m} 
- \boldsymbol{z}_{u}^{T} \boldsymbol{J}_{\boldsymbol{u},\boldsymbol{z}} \boldsymbol{z}
\right\}
\end{multline}

The next challenge is obtaining $\boldsymbol{J_{u,z}}$. Note that by adopting the Gaussian variant of SimplE, the likelihood factor between the user belief distribution $\boldsymbol{z}_{u}$ and item distribution $\boldsymbol{z}_{m}$ becomes $\exp \{-\langle \boldsymbol{z}_{u}, \boldsymbol{r}, \boldsymbol{z}_{m} \rangle \}$ where $\boldsymbol{r}$ is the embedding vector of the \textit{likes} relation --- this is log-bilinear in $\boldsymbol{z}_{u}$ and $\boldsymbol{z}_{m}$ and would appear to stymie closed-form Gaussian belief propagation. 
Serendipitously, we can rewrite $\langle \boldsymbol{z_u}, \boldsymbol{r}, \boldsymbol{z_m} \rangle$ as $\boldsymbol{u}^{T} \boldsymbol{D_{r}} \boldsymbol{z}$, where $\boldsymbol{D_{r}}$ is derived by reshaping $r$ as a diagonal matrix. Hence, we have $\boldsymbol{J}_{\boldsymbol{u},\boldsymbol{z}} = \boldsymbol{D}_{r}$. Next,
to derive $p(\boldsymbol{z}_{u}|\boldsymbol{z_d})$, the posterior user belief distribution, we marginalize the joint probability of the user belief and item distribution over $\boldsymbol{z}_{m}$:
\begin{align}
        p(\boldsymbol{z}_{u}|\boldsymbol{z}_{d}) = & \int {p(\boldsymbol{z}_{u}, \boldsymbol{z}_{m}|\boldsymbol{z}_{d})} d\boldsymbol{z}_{m} \propto \exp\left\{{-\frac{1}{2}}\boldsymbol{z}_{u}^{T} \boldsymbol{J}_{\boldsymbol{u}} \boldsymbol{z}_{u} + \boldsymbol{h}_{\boldsymbol{u}}^{T} \boldsymbol{z}_{u} \right\}  \\ & \times \int { \exp \left\{ - \frac{1}{2}\boldsymbol{z}_{m}^{T} \boldsymbol{J}_{\boldsymbol{z}} \boldsymbol{z}_{m} + \boldsymbol{h}_{\boldsymbol{z}}^{T} \boldsymbol{z}_{m} - 
        \boldsymbol{z}_{u}^{T} \boldsymbol{J_{u,z}} \boldsymbol{z}_{m} \right\}d\boldsymbol{z}_{m}} \nonumber
\end{align}
In order to calculate the latter integral, we first write the exponentiated term in matrix form to obtain:
\begin{align}
p(\boldsymbol{z_u}|\boldsymbol{z_d}) &\propto  \exp\left\{{-\frac{1}{2}}\boldsymbol{z}_{u}^{T} \boldsymbol{J}_{\boldsymbol{u}} \boldsymbol{z}_{u} + \boldsymbol{h}_{\boldsymbol{u}}^{T} \boldsymbol{z}_{u} \right\} \times  \\ & 
        \int{ \exp \left\{ -\frac{1}{2} \begin{pmatrix}
             \boldsymbol{z}_{u}\\
             \boldsymbol{z}_{m}
         \end{pmatrix} ^{T} \begin{pmatrix}
             \boldsymbol{0} &  \boldsymbol{J}_{\boldsymbol{u},\boldsymbol{z}} \\
             \boldsymbol{J}_{\boldsymbol{u},\boldsymbol{z}} & \boldsymbol{J}_{\boldsymbol{z}}
         \end{pmatrix} \begin{pmatrix}
             \boldsymbol{z}_{u} \\
             \boldsymbol{z}_{m}
         \end{pmatrix}  +  \begin{pmatrix}
             \boldsymbol{0} \\
             \boldsymbol{h}^{T}_{\boldsymbol{z}}
         \end{pmatrix}  \begin{pmatrix}
             \boldsymbol{z}_{u} \\
             \boldsymbol{z}_{m}
         \end{pmatrix} \right\} d\boldsymbol{z}_{m}} \nonumber 
\end{align}
Now, observing that the term inside the integral is \emph{exactly} in the form of a jointly Gaussian distribution, we can obtain its marginal probability distribution using the Schur complement. Thus, we have $p(\boldsymbol{z}_{u}) \sim \mathcal{N}^{-1}(\hat{\boldsymbol{h}_{\boldsymbol{u}}}, \hat{\boldsymbol{J}_{\boldsymbol{u}}})$, where:
\begin{equation}
    \hat{\boldsymbol{h}}_{\boldsymbol{u}} = \boldsymbol{h}_{\boldsymbol{u}} - \boldsymbol{J}_{\boldsymbol{u},\boldsymbol{z}} ( \boldsymbol{J}_{\boldsymbol{z}})^{-1} (\boldsymbol{h}_{\boldsymbol{z}})
\end{equation}
\begin{equation}
    \hat{\boldsymbol{J}}_{\boldsymbol{u}} = \boldsymbol{J}_{\boldsymbol{u}} - \boldsymbol{J}_{\boldsymbol{u},\boldsymbol{z}} ( \boldsymbol{J}_{\boldsymbol{z}})^{-1} \boldsymbol{J}_{\boldsymbol{u},\boldsymbol{z}}
\end{equation}
To summarize, 
while the use of a tensor-based likelihood introduced an unusual log-bilinear form and the need to marginalize over the latent item distribution induced by the KG critiques, we have shown that we can manipulate all necessary quantities in Gaussian form.
In this way, we can perform a closed-form Gaussian user belief update w.r.t.\ an item distribution inferred by indirect KG properties.
\begin{figure*}
\begin{minipage}[t]{0.48\linewidth}

    \includegraphics[height=3.5cm]{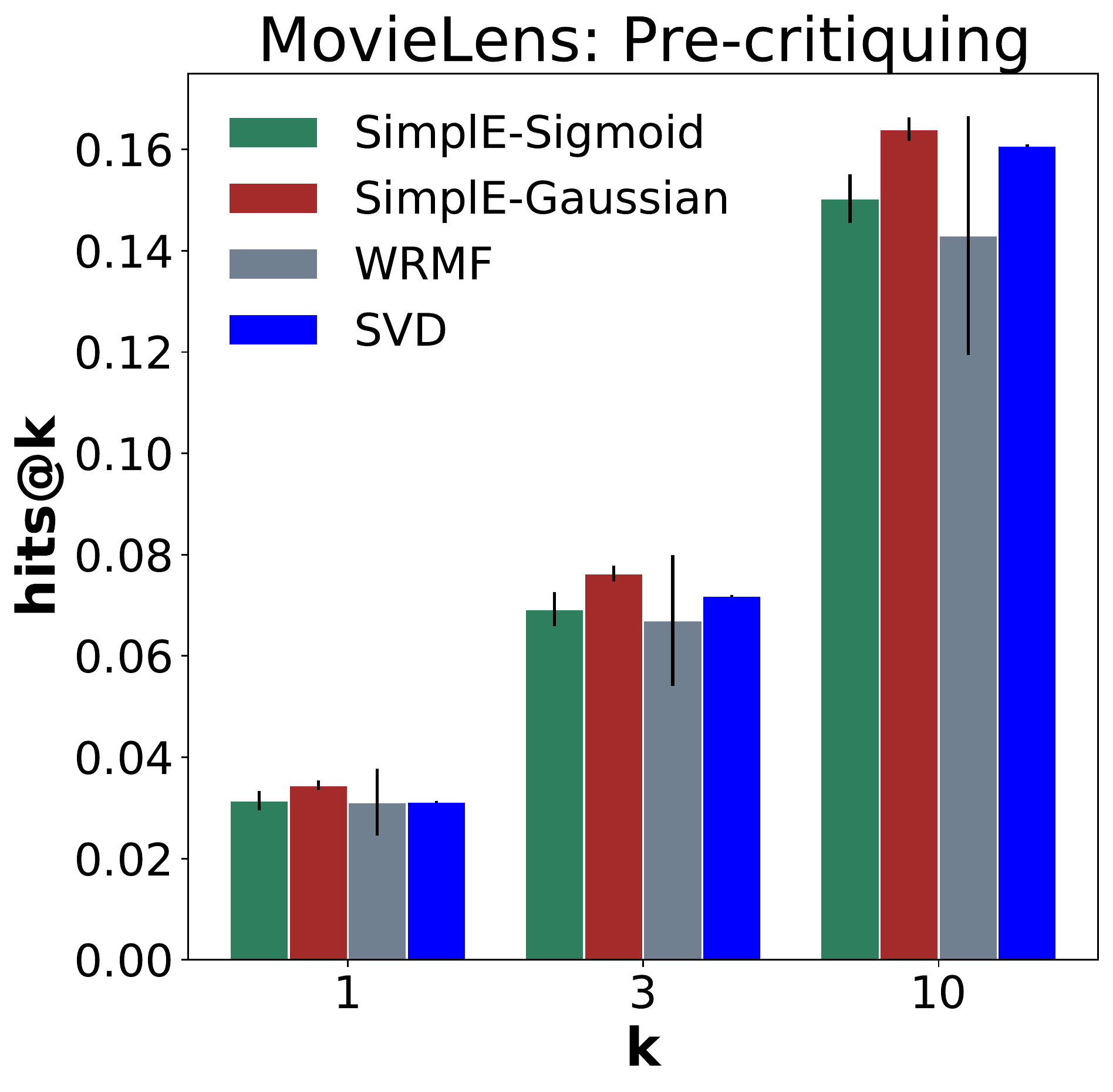}
    \hspace{28pt}
     \includegraphics[height=3.5cm]{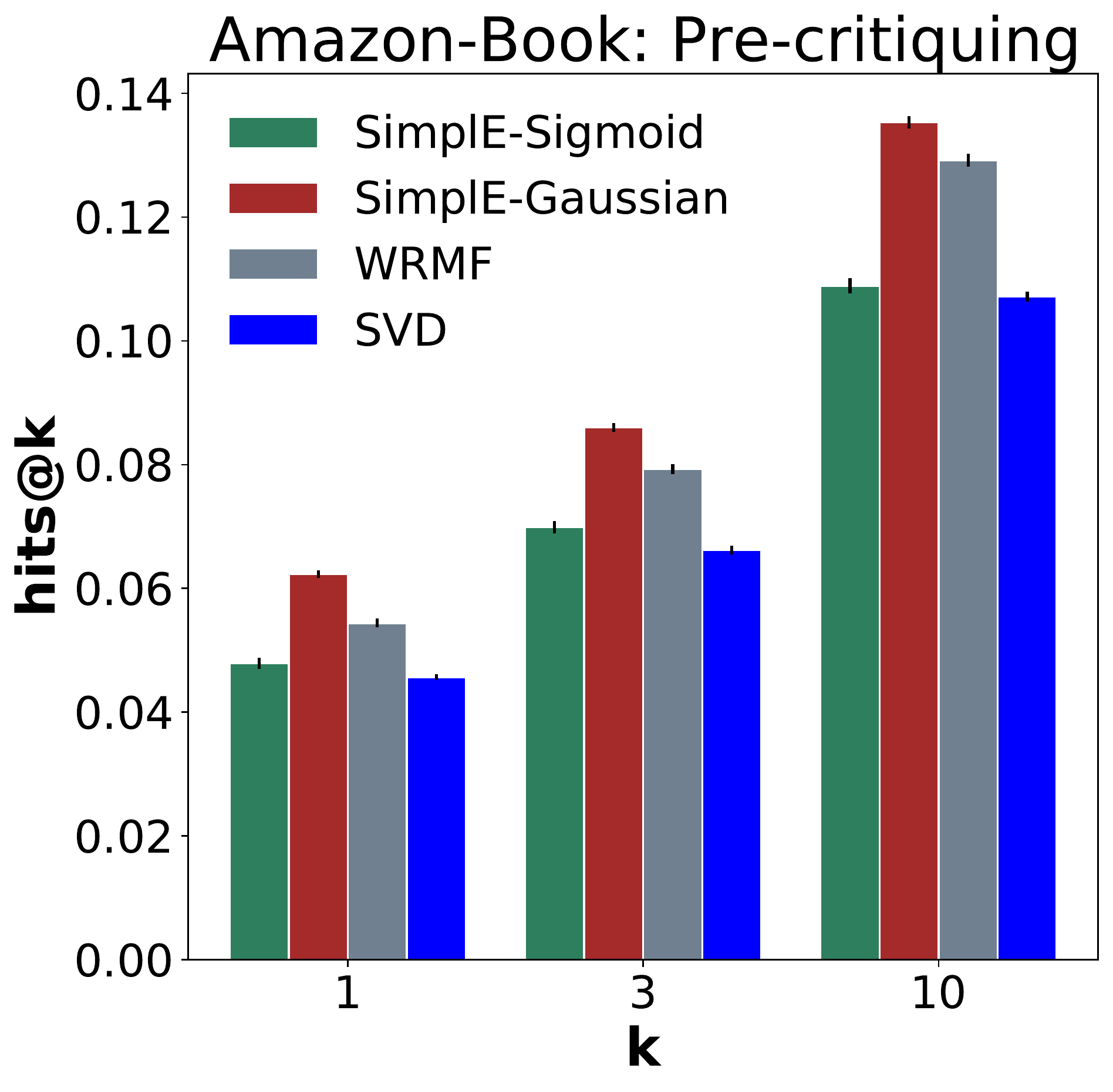}
    \caption{Initial performance of tensor factorization-based recommender compared to classical recommendation models}
    \label{fig:RQ1}
    \end{minipage}
    \hspace{2pt}
    \begin{minipage}[t]{0.48\linewidth}
    \hspace{2pt}
    \includegraphics[height=3.5cm]{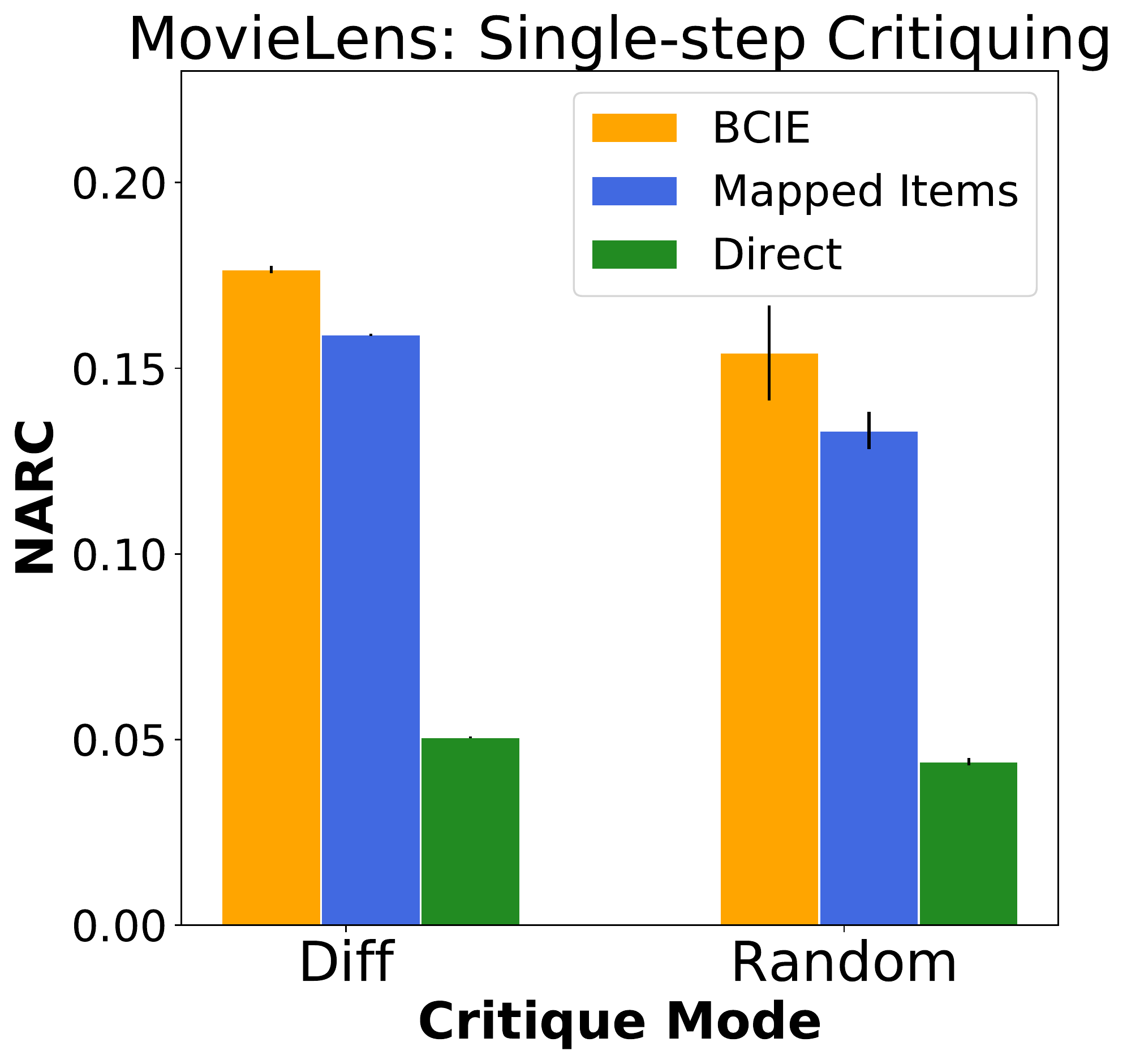}
    \hspace{14pt}
     \includegraphics[height=3.5cm]{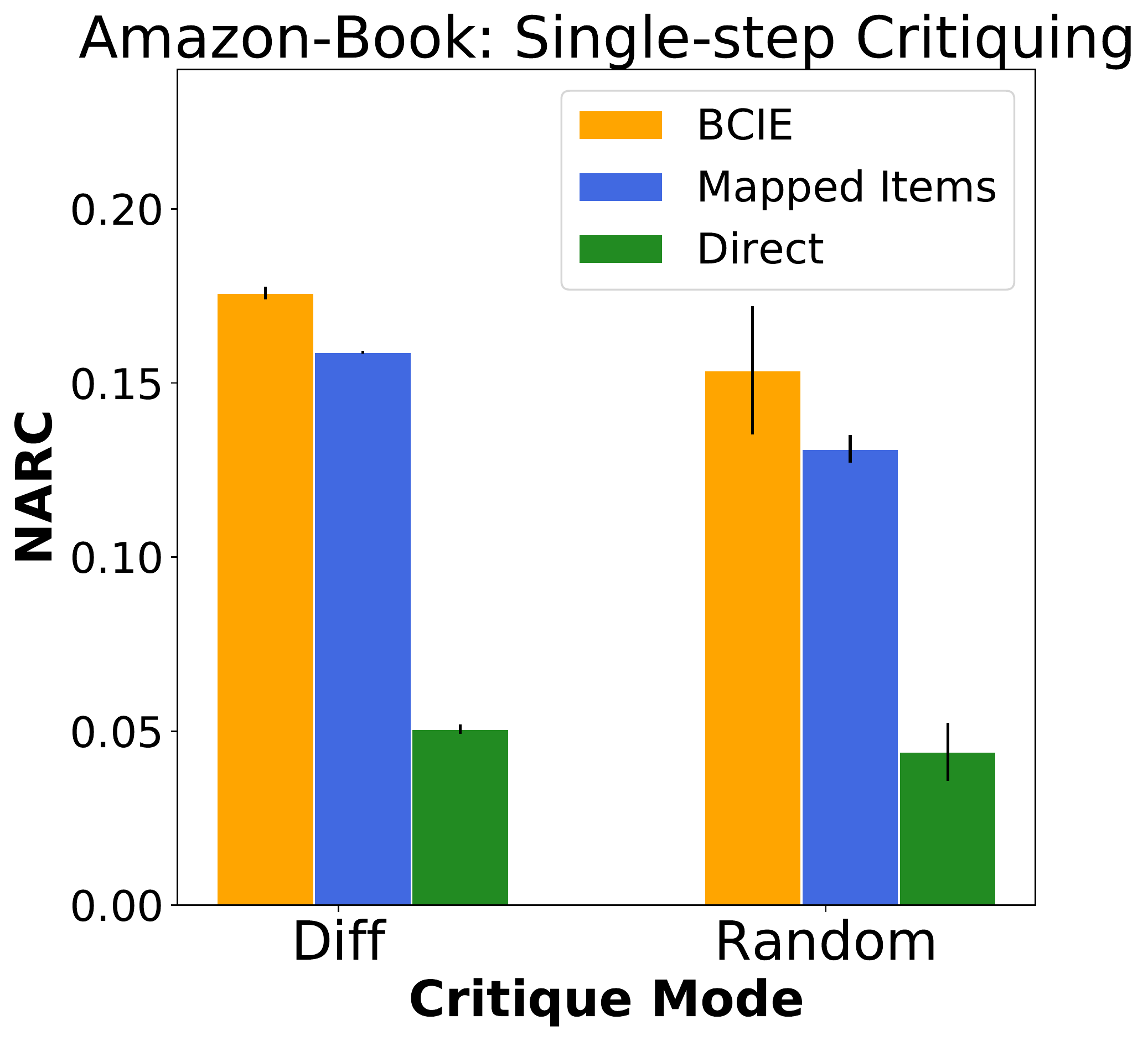}
        \caption{Single-step critiquing performance of BCIE compared to the baselines under two critique modes}
    \label{fig:RQ2}
    \vspace{1mm}
\end{minipage}

\end{figure*}
\begin{figure*}[!t]
    \centering
\centering
\begin{subfigure}{0.247\textwidth}
\includegraphics[width=0.92\textwidth]{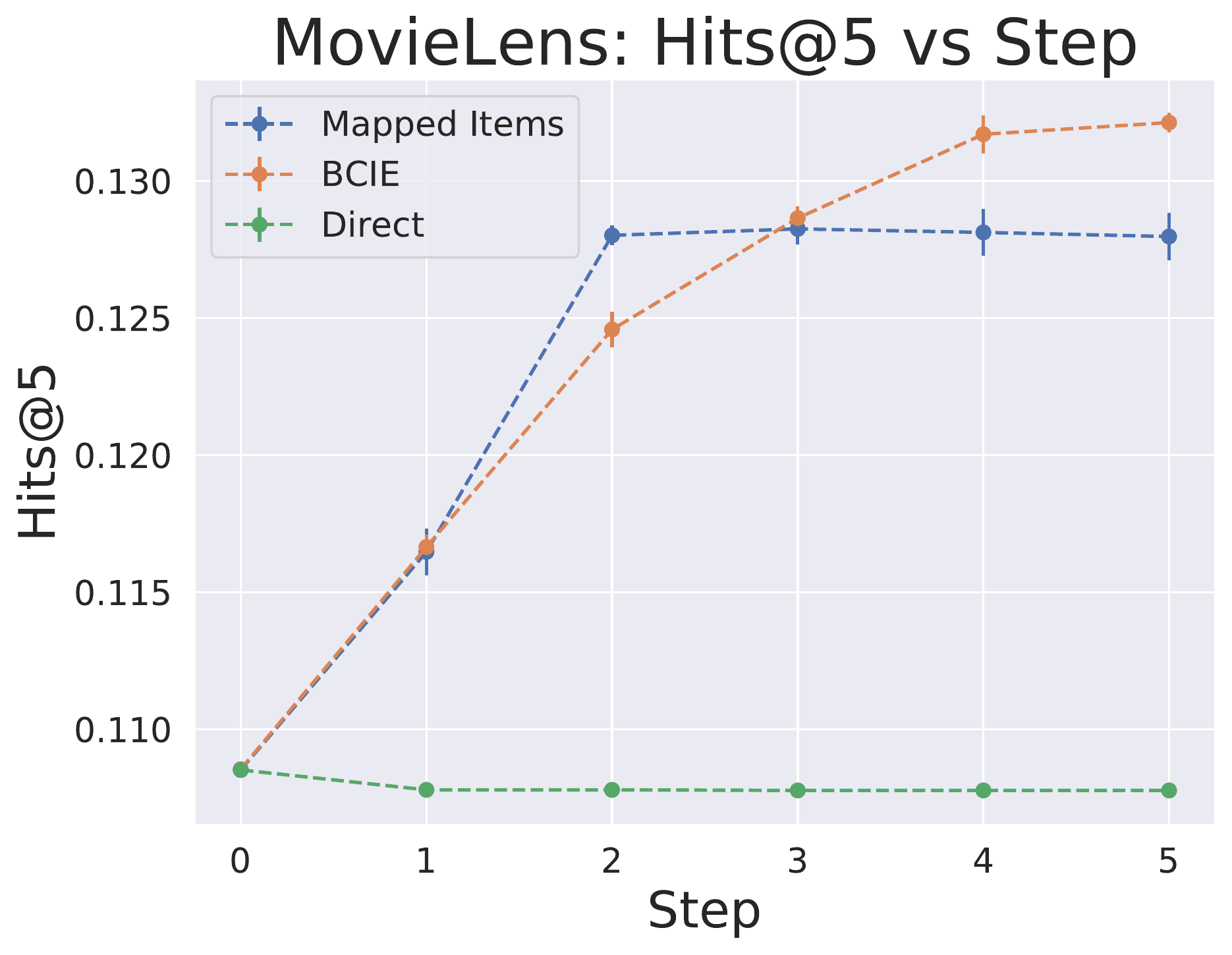}
\end{subfigure}
\begin{subfigure}{0.247\textwidth}
\includegraphics[width=0.92\textwidth]{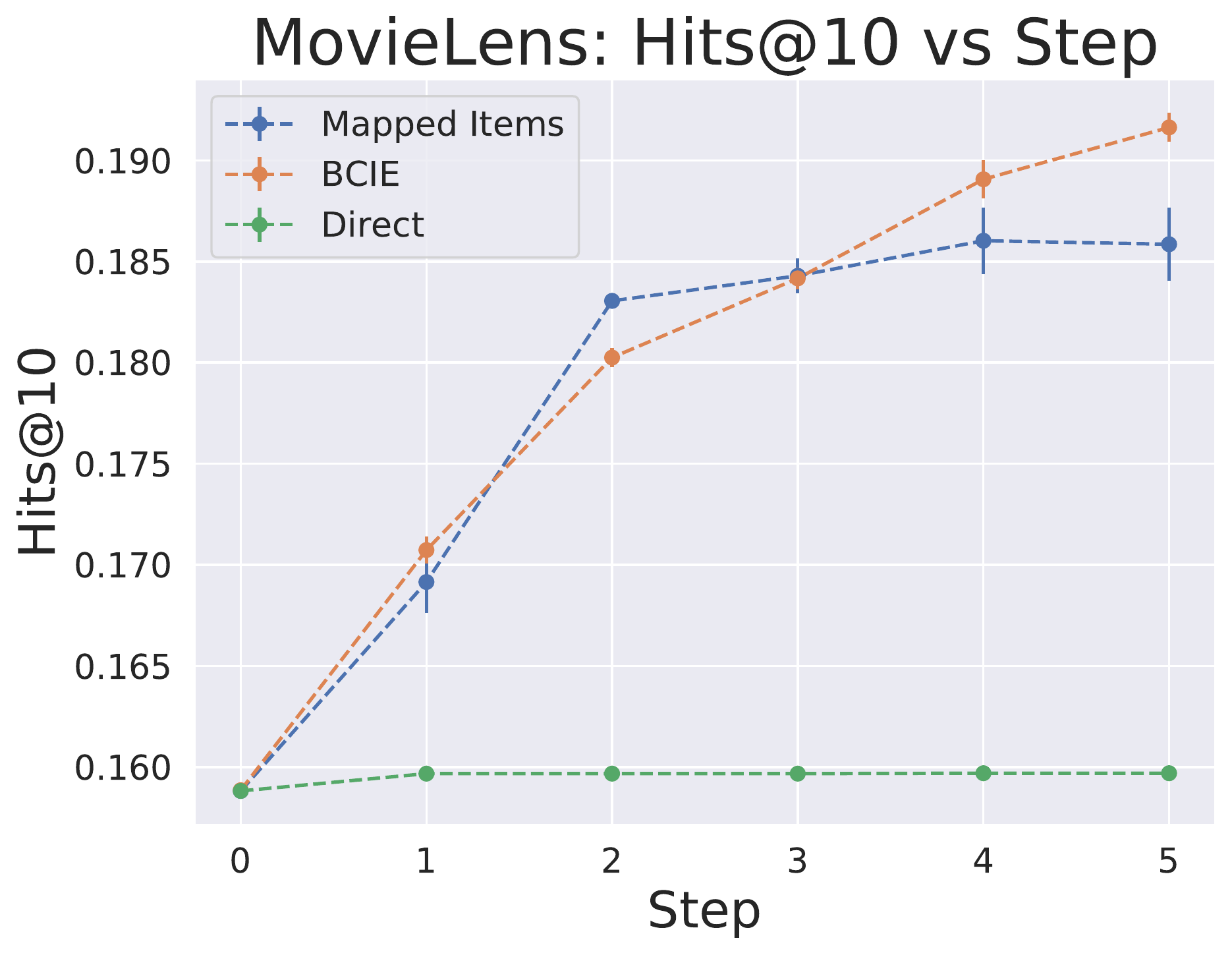}
\end{subfigure}
\begin{subfigure}{0.247\textwidth}
\includegraphics[width=0.92\textwidth]{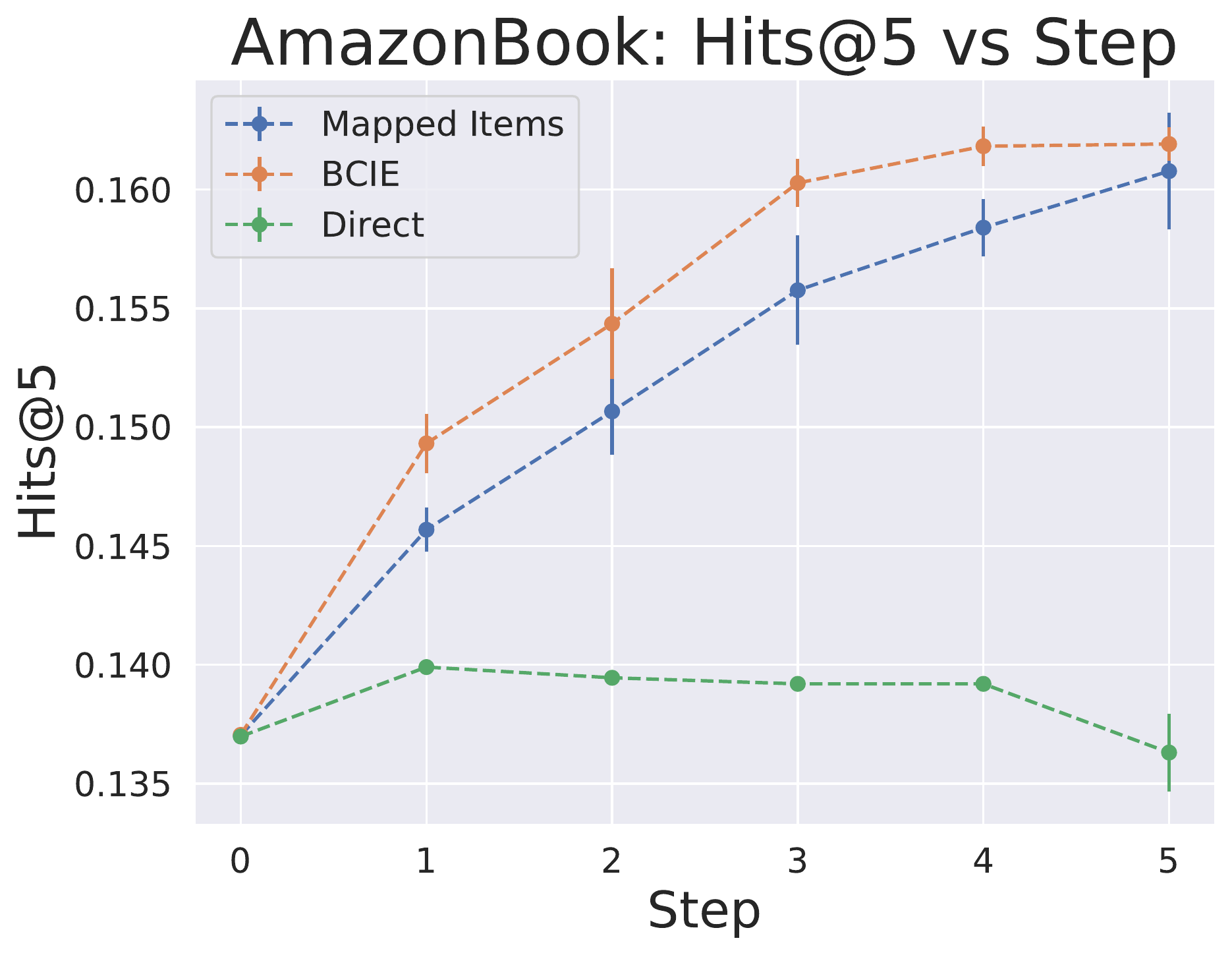}
\end{subfigure}
\begin{subfigure}{0.247\textwidth}
\includegraphics[width=0.92\textwidth]{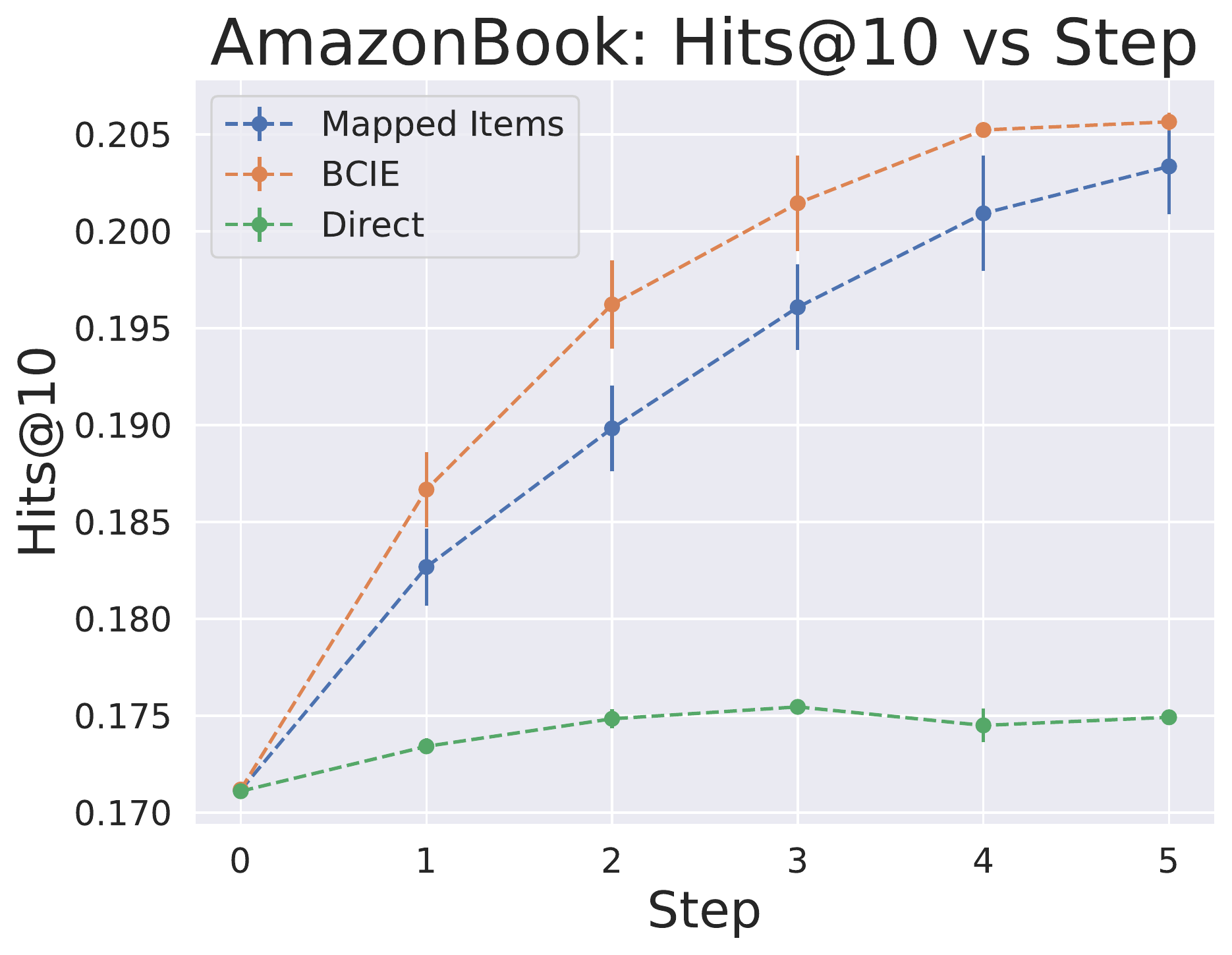}
\end{subfigure}

\vspace{-1mm}
\caption{Comparison of change in hit rate @\{5,10\} during the critiquing sessions averaged over 3 independent runs.}
\label{rq3}
 \end{figure*}
 \begin{table}
  \caption{Summary of Dataset Statistics}
  \label{tab:data}
  \begin{tabular}{ccccc}
    \toprule
    Name & \#Users & \#Items & \#KG entities & \#KG relations \\
    \midrule
    MovieLens & 61,715 & 16,826 & 56,789 & 46 \\
    AmazonBook & 74,026 & 21,081 & 79,682 & 38 \\
  \bottomrule
\end{tabular}
\end{table}
\section{Experiments and Evaluation}
\label{sec:dataset}

\subsection{Datasets}
We evaluate BCIE\footnote{\url{https://github.com/atoroghi/BCIE}} on two of the most popular recommendation datasets, MovieLens 20M \footnote{\url{https://grouplens.org/datasets/movielens/20m/}} and Amazon-Book \footnote{\url{http://jmcauley.ucsd.edu/data/amazon/}}, and acquire facts about their items from Freebase KG \citep{freebase}. 
We consider ratings greater than 3.5 to indicate that the user likes an item and extract facts about items from Freebase using entity matching data from \cite{zhao2019kb4rec}.
Also, since we conduct 5 steps of critiquing, we only keep items with at least 5 KG facts to enable selection of non-repetitive critiques.
Table \ref{tab:data} shows dataset statistics.

\subsection{Baselines}
As prior KG-enhanced recommendation studies have not considered the conversational setting and previous critiquing 
works do not handle knowledge-based indirect evidence, we propose two comparison baselines, namely \textit{'Mapped items'} and \textit{'Direct'}. \textit{Mapped items} is a heuristic method that maps each critique to a maximum of 10 relevant items from the KG and uses them for user belief updating. For example,
for the critique \textit{"I prefer movies like Nolan's works"}, movies that are \textit{directed by, Nolan} are mined from the KG and used as examples of the user's interests. Notably, \textit{Mapped items} is anticipated to be a strong baseline, as the ground truth for many critiques, especially those referring to less popular entities is likely among the few extracted items, 
and thus, revealed to the recommender. 
\textit{Direct} is an ablation of BCIE that treats indirect evidence of the user's interest as direct by omitting the item distribution and directly attributing the critique fact to the user belief.

\subsection{RQ1: Pre-critiquing recommendation performance.}
Although the pre-critiquing performance is not our main focus, it can indicate the quality of the trained embeddings and provide a baseline for comparing the efficiency of critiquing in improving the recommendations. We use two classical recommendation models, namely WRMF \citep{wrmf} and SVD \citep{svdmodelbased},  as the baselines. For all experiments, hyperparameter tuning is performed using nested cross-validation with one inner loop validation fold and five test folds in the outer loop. Results of these experiments are provided in Figure \ref{fig:RQ1} that show the proposed Gaussian variant of the tensor factorization-based recommender outperforms the classical models and the performance of the original SimplE model is on par with that of these baselines.

\subsection{RQ2: Effect of Single-step Critiquing.}
As a proof-of-concept, we investigate BCIE's efficacy in single-step critiquing. Following a positive critique,
we anticipate observing a reduction in the average rank of items satisfying the fact. To monitor this impact, we utilize the Normalized Average Rank Change (NARC) metric defined as
\begin{equation}
    NARC = \frac{AR_{pre}(I^{f}) - AR_{post}(I^{f})}{AR_{pre}(I^{f})}
\end{equation}
where $AR_{pre}(I^{f})$ and $AR_{post}(I^{f})$ denote the Average Rank of items fulfilling the critique fact $f$ before and after
updating the user belief, respectively. This experiment is conducted under two critique selection modes: \textit{diff}, where the user critiques a fact related to ground truth that deviates the most from the recommended items facts, and \textit{random}, where the critique fact is randomly selected. In all critiquing experiments, the only hyperparameters are the initial precision matrices which are optimized for all methods. The experimental results for both datasets are presented in Figure \ref{fig:RQ2} 
demonstrating that BCIE is more proficient in capturing the user's interests from their critiques and reflecting them in reducing the rank of the relevant items by more than 15 \%.

\subsection{RQ3: Multi-step Critiquing Performance.}
In this experiment, for each $(u, likes, i)$ triple in the test set, item $i$ is once considered as the ground truth target of recommendation for user $u$. At each critiquing session, the top 10 ranked items 
together with 
their KG critiques are presented to the user, and following \textit{diff} critique selection mode, a critique is selected. Experimental results comparing BCIE with the baselines are presented in Figure \ref{rq3}. The hit rates at 5 and 10 reveal that BCIE often performs superior to the \textit{Mapped Items} and \textit{Direct} baseline
, demonstrating the effectiveness of BCIE's Bayesian update of the user belief in 
accurately
capturing the user's preferences and directing the top recommendations towards the items of interest. The initial hit rate increase of the \textit{Mapped items} in the MovieLens dataset is observed to be caused by the fact that, in several cases, the ground truth is among the few items used for updating the user belief from the first critique. However, since further updating with the same evidence does not contribute new information, the performance plateaus. Also, the comparison with the direct baseline highlights the importance of incorporating item distribution in the framework.
It is noteworthy that treating indirect evidence feedback (e.g., \textit{"I want movies related to Vietnam war"}), as direct evidence of the user's interest (e.g., \textit{"User likes Vietnam war"}) significantly alters the semantics.
\section{Conclusion}
We proposed BCIE, a novel Bayesian framework that enables incorporating rich contextual and KG-based relational side information into critique-based 
recommender systems to enhance their expressivity by allowing for knowledge-based indirect feedback. Experimental results exhibit the efficacy of BCIE in initial recommendation and its ability to capture user preferences by reasoning over the indirect evidence observed through sequential critiquing. This promising performance combined with the computational efficiency of its closed-form posterior user belief update opens up possibilities for versatile future extensions to address more complex and natural types of user critiquing feedback.


\bibliographystyle{ACM-Reference-Format}
\balance
\bibliography{sample-base}


\end{document}